# Copying Versus Randomization in Lempel-Ziv Music Synthesis

Nadav Mishan
Department of Electrical Engineering
Tel Aviv University
Tel Aviv, Israel
nadavmishan@mail.tau.ac.il

Ram Zamir
Department of Electrical Engineering
Tel Aviv University
Tel Aviv, Israel
https://cris.iucc.ac.il/en/persons/ram-zamir

***Abstract*— We utilize Lempel-Ziv universal compression for music note generation. We control the algorithm's tendency to over-copy or under-copy training data by manipulating the average sequence length saved in the dictionary.**

## I. Introduction

The field of generative artificial intelligence (GenAI) is currently dominated by diffusion models, transformer-based models, and generative adversarial networks (GANs) [1]. While effective, these methods demand substantial computational resources and obscure the logic of their creation process. Universal Compression (UC) models, conversely, operate efficiently in "real-time". This computational advantage, combined with the ability of algorithms like Lempel-Ziv (LZ) [2] to learn statistical models [3] for data compression, presents a promising alternative path for GenAI. Specifically, the LZ algorithm has been explored for music generation in prior works such as Dubnov and Assayag [4] and recently by Ding et al. [5].

The fundamental principle of LZ compression is the detection and reuse of repeated sequences, which poses a significant challenge for content generation. In the musical domain, new compositions are typically inspired by existing works rather than reproduced from them. However, the standard LZ algorithm is designed to maximize compression efficiency by copying long, contiguous segments of the source data. Consequently, the resulting output often consists of verbatim excerpts from the training material, rather than exhibiting the originality and variation expected of creative composition [6].

In this work, we propose a novel method to address this limitation by introducing controls to balance the trade-off between exact copying (over-copy) and randomization (under-copy). We propose two methods of controlling average sequence length: limiting the training data per tree to decrease average sequence length, and a sequence extension technique that extends branch depth to increase average sequence length.

## II. Methodology

In this section we describe the learning and generation process of the algorithm. An overview of the process is shown in Fig. 1:

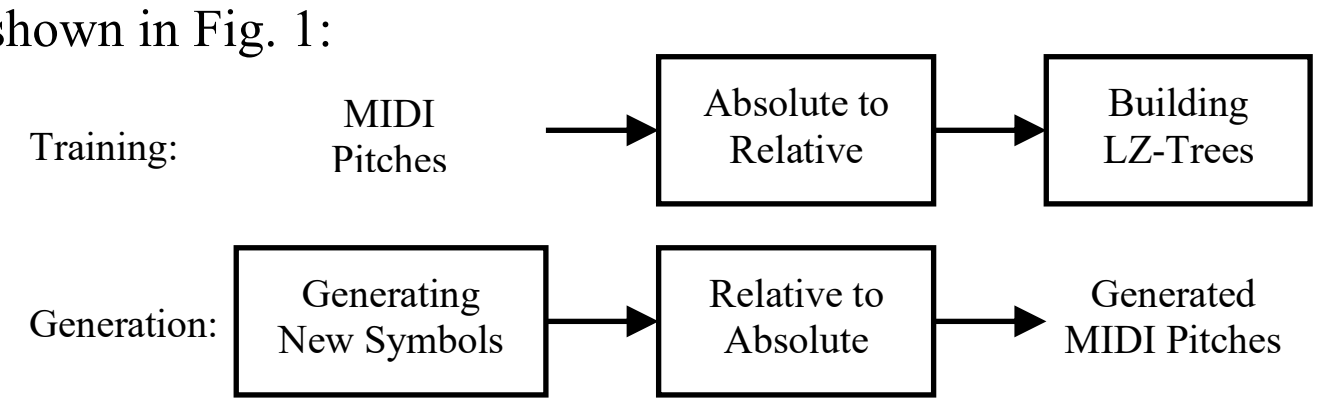

*Fig. 1. Block Diagram of the Training -Generation Process*

Code implementation in Python is available at the GitHub repository [7].

### A. Training Data and Preprocessing

We use Google-magenta's MAESTRO [8] database which contains about 200 hours of annotated piano performances. Specifically, only MIDI [9] pitch data is used, ignoring velocity, and Note-On/Off events.

To accentuate repeating patterns across different compositions, which vary in octave and key, we apply a delta transform to the MIDI data. This transforms the sequence of absolute note values into a sequence of relative intervals (differences between consecutive notes). Consequently, the LZ algorithm processes the interval structure of the music rather than specific pitch frequencies.

### B. Building the LZ tree

The algorithm uses the LZ78 [10] branch of the LZ family, where seen sequences are kept in a dictionary for later reference. In our work we symbolize the learned dictionary as a tree. The algorithm starts with a pointer to the root of an empty tree and in each iteration, it checks if the next symbol in the data exists in the node's children:

*a) The Symbol Exists: (Seen Sequence)*

It raises the frequency parameter of the child node by one and moves the pointer to the node.

*b) The Symbol Doesn't Exist: (New Sequence)*

It creates a new child for the current node with the current symbol (extending the sequence) and moves the pointer back to the root.

At this stage we break from the regular LZ practice and explore a method of creating longer sequences. Instead of going back to the root after extending the sequence by one symbol (step), We propose a parameter - "steps" that sets the algorithm to go back to root after a certain number of steps taken. Thus, *increasing* the average sequence length.

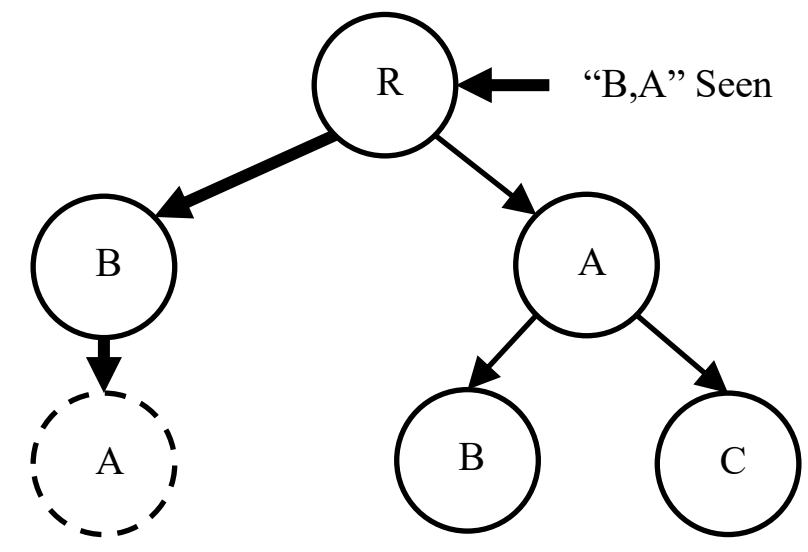

*Fig. 2. Example of LZ-Tree Building*

Another way we control tree depth is by limiting the amount of data. As described by Cover and Thomas in thm. 13.5.2 [2], when a stationary ergodic process is compressed by LZ the number of phrases (sequences) in the dictionary is described by:

$$\limsup_{n\to\infty} \frac{c(n)\log_2 c(n)}{n} \le H(x) \quad (1)$$

Where $H(x)$ is the entropy rate of the process, $c(n)$ is the number of phrases in a distinct parsing of a sample length $n$ from this process.

$c(n)$ can also be estimated by Eq. 6 [10]:

$$c(n) < \frac{n}{\log_2 n} \log_2 \alpha \quad (2)$$

Where $\alpha$ is a finite input symbol alphabet. Combining the equations, we create a heuristic for the average length of a sequence, $L_s$:

$$L_S \triangleq \frac{n}{c(n)} \approx \frac{\log_2 n}{H(x)} \quad (3)$$

This heuristic shows us that the average sequence length grows with $n$. To control this behavior, we chose a tree ensemble approach: A set of trees is created, and each tree is given a subset of the data to train on this *decreasing* $L_s$.

### C. Generation

To generate a new sequence the algorithm goes through the following steps:

I. A tree is uniformly picked from the set of trees.
II. The next symbol to output is chosen probabilistically, based on the frequency each symbol has been seen before.
III. Repeat step II until a leaf node is reached.
IV. A new tree from the set is chosen.
V. Repeat steps I-IV until the number of symbols chosen are generated.

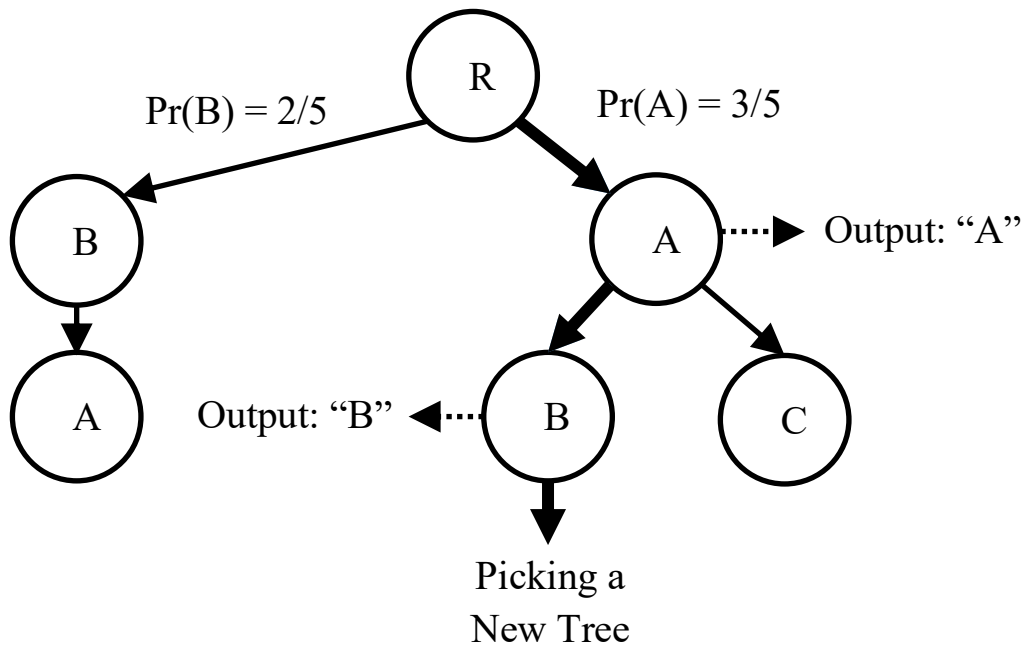

*Fig. 3. Example of the Generation Process*

## III. Experiments

As mentioned in section II.B. we tried two different methods to control $L_S$:

### A. Limiting Training Data

As Eq. (3) suggests, limiting each tree's data controls the average sequence length:

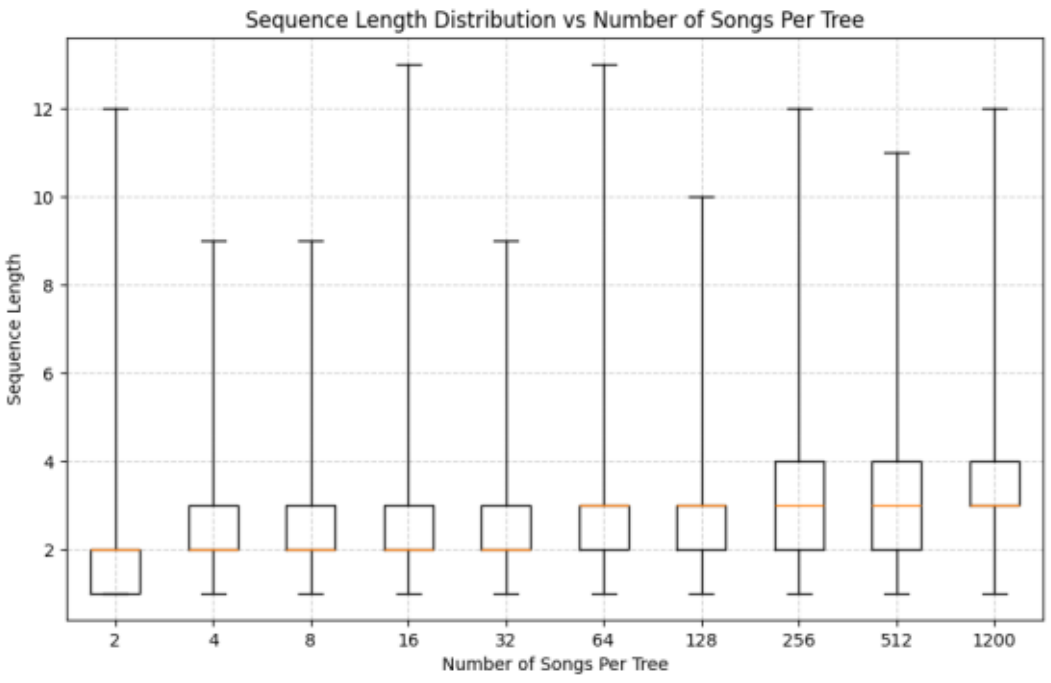

*Fig. 4. Average Sequence Length as a Function of the Number of Songs the Tree Trained on*

In Fig. 4 we can see an empirical result for Eq. (3). As we decrease the amount of data each tree receives, $L_s$ decreases.

### B. Choosing the Number of Steps

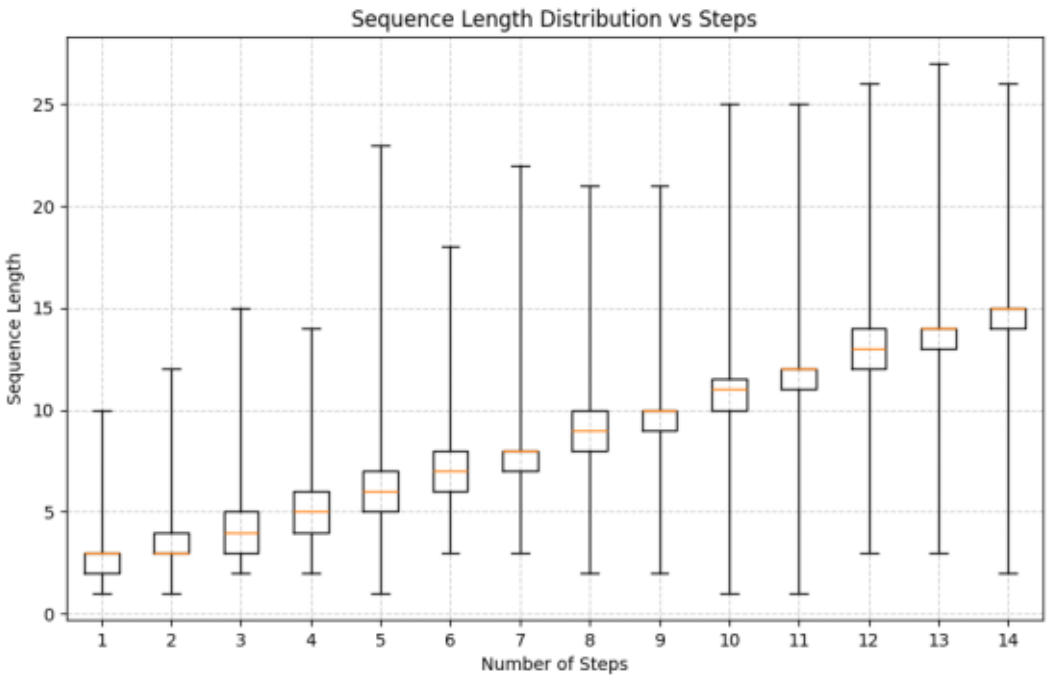

*Fig. 5. Average Sequence Length as a Function of Steps*

In Fig. 5 shows that, as intended, increasing "steps" increases $L_s$ (3).

### C. Qualitative Evaluation

Examples of songs created by the algorithm, with different parameters, are available at the GitHub repository [7]. Listening to songs uncovers problems in our algorithm:

#### 1) Rhytmic Ridgidity

As the training process omitted rhythm, the generated output exhibits a strict bit. The resulting lack of timing deviations contributes to a perceptual "robotic" quality, highlighting the necessity of encoding expressive timing features in future iterations.

#### 2) Structure Coherence vs. Overfitting

We observe a direct correlation between $L_s$ and musical coherence. Samples generated with larger $L_s$ values exhibit established chord progressions and consistent tone. However, as noted in Section I, this coherence often stems from the algorithm copying verbatim sections of the training data (overfitting). We propose using methods for plagiarism detection [6] to test the song's uniqueness.

## IV. Conclusions

As we see from the results, Our algorithm can control the balance between copying and randomization with low complexity. In the future we would like to utilize different UC algorithms and improve the structure of the songs through pre-processing and a different generation stage.